# Brillouin microscopy – a revolutionary tool for mechanobiology?

Robert Prevedel[1-3,*], Alba Diz-Muñoz[1,*], Giancarlo Ruocco[4,5] and Giuseppe Antonacci[4,6]

[1] Cell Biology and Biophysics Unit, European Molecular Biology Laboratory, Heidelberg, Germany
[2] Developmental Biology Unit, European Molecular Biology Laboratory, Heidelberg, Germany
[3] Epigenetics and Neurobiology Unit, European Molecular Biology Laboratory, Monterotondo, Italy
[4] Center for Life Nano Science, Istituto Italiano di Tecnologia, Rome, Italy
[5] Department of Physics, Sapienza University of Rome, Rome, Italy
[6] Photonics Research Group, Ghent University – IMEC, Ghent, Belgium

*: Correspondence should be addressed to R.P. (prevedel@embl.de) or A.D.-M. (diz@embl.de)

**The role and importance of mechanical properties of cells and tissues in cellular function, development as well as disease has widely been acknowledged, however standard techniques currently used to assess them exhibit intrinsic limitations. Recently, a new type of optical elastography, namely Brillouin microscopy, has emerged as a non-destructive, label- and contact-free method which can probe the viscoelastic properties of biological samples with diffraction-limited resolution in 3D. This has led to increased attention amongst the biological and medical research communities, but also to debates about the interpretation and relevance of the measured physical quantities. Here, we review this emerging technology by describing the underlying biophysical principles and discussing the interpretation of Brillouin spectra arising from heterogeneous biological matter. We further elaborate on the technique's limitations as well as its potential for new insights in biology in order to guide interested researchers from various fields.**

## Introduction

Across spatial scales, the mechanical properties of cells and tissues are important, as they play intricate roles in determining biological function[1–3]. On the cellular scale, elastic and viscous properties have emerged as key parameter regulating cell differentiation and migration, and in determining how cells respond to physical forces and their environment. Changes in cell stiffness, for example, have been associated with immune and epithelial tumour aggressiveness[4,5] and the level of stemness in limbal stem cells[6]. The elasticity of the extracellular environment, on the other hand, directs lineage specification in stem cells[7], drives tumour progression[8,9] and regulates cadherin-dependent collective migration[10]. At a tissue level, mechanical properties of tissues are dominant drivers of morphogenesis and multicellular organization[11], and are thought to be imperative in the onset and progression of many diseases, such as eye disease[12,13], cancer[14] or atherosclerosis[15].

In recent years, substantial progress has been made towards understanding how these mechanical cues, either extrinsically induced by the cellular microenvironment or intracellularly generated, can be transduced into biochemical signals that regulate cell proliferation, migration and tissue dynamics[16–21]. While molecular components can routinely be visualized *in-situ* with powerful tools such as fluorescence microscopy, current approaches to measure cell mechanical properties *in-vivo* have important limitations[22–24] and assessing the mechanical properties of living cells and tissues with similar spatial-temporal resolution in a non-invasive fashion has long been an open challenge (see **Box 1**). In addition, these approaches can only measure the quasi-static Young's modulus, which in turn has been conceptually and purely conventionally linked by biologists to the meaning of *stiffness*.

---

**Box 1 | Current techniques to infer mechanical properties in biology**

Existing methods to measure viscoelastic properties either require contact forces or lack appropriate subcellular resolution in 3D. Atomic force microscopy (AFM), the current gold standard in the field of mechanobiology, involves the application of nano-indenters to surfaces, e.g. the cellular membrane, to measure the quasi-static Young's modulus $E$ from the deflection of a cantilever. While this can provide high transverse spatial resolution on the nanometer scales[25], measurement are averaged along the contact (axial) direction and rely heavily on mechanical models to extract $E$ values. Other approaches to measure elasticity include micropillar deformation[26], micropipette aspiration[27], deformability cytometry[28], magnetic twisting cytometry[29] and optical tweezers[30]. Furthermore, microrheology based on optical tweezing, can measure viscosity on the micron length scale[31,32]. However, these methods either require direct contact to the cells of interest, rely on the introduction of foreign particles or do not work in multi-cellular situations. Moreover, a recent study highlighted the intrinsic variability of those techniques when directly compared among each other[24]. Other optical approaches, such as optical coherence elastography (OCE), requires external contact forces or ultrasound fields to measure tissue displacements[33]. While OCE enables rapid three-dimensional imaging, current implementations do not achieve cellular resolution, a limitation shared with other noninvasive techniques, such as ultrasound[34] and magnetic resonance imaging[35].

---

Brillouin spectroscopy was originally developed[36] and widely utilized in the context of material science[37,38], where it became a powerful and well-established technique for the investigation of condensed matter. It provides an alternative and complimentary assessment of material elasticity and viscosity through measurement of the longitudinal modulus in the GHz frequency range. The introduction of Brillouin spectroscopy to biology[39] and its combination with scanning confocal microscopy has further stimulated the recently booming field of biomechanics, enabling a new way to directly 'image' viscoelastic properties of living matter in 3D and in a noncontact, label-free and high-resolution fashion.

Brillouin microscopy has enabled a wide range of applications since its first demonstration a decade ago, including the investigation of intracellular biomechanics in whole living cells[40–42], the analysis of liquid-to-solid phase transitions in individual subcellular structures[43], the biomechanical assessment of regenerating tissues in living zebrafish[44] as well as the 3D mapping of the lens cornea biomechanics both *ex-vivo*[45,46] and *in-vivo*[47]. It furthermore holds great promises for the early diagnosis of diseases such as atherosclerosis[48], cancer[49], keratoconus[50], meningitis[51], Alzheimer's[52] and amyotrophic lateral sclerosis[43]. Yet, despite its seemingly revolutionary advantages, the biophysical interpretation of Brillouin microscopy measurements is still heavily debated[53,54] and their relevance for many current questions in biomechanics is still unclear. In this Review, we introduce the physical principles underlying the Brillouin scattering process, and discuss how to infer mechanical properties from the spectra in light of both the complex dynamics in heterogeneous biological matter and the instrument's limitations. We further discuss how Brillouin measurements relate to cell- and tissue-mechanical properties, highlight differences to viscoelastic measurements done with existing techniques in the field and describe open challenges in translating recent technical advances into new biological insight.

**Underlying physical principles**

First reported in 1922[36], Brillouin light scattering is an inelastic process arising from the interaction of light with spontaneous, thermally induced density fluctuations. These can be considered as microscopic acoustic waves (with wavelength $\Lambda$ and period $T$ related by $\Lambda = VT$, where $V$ is the medium's sound velocity), and are often called phonons (see **Fig.1** and **Box 2**).

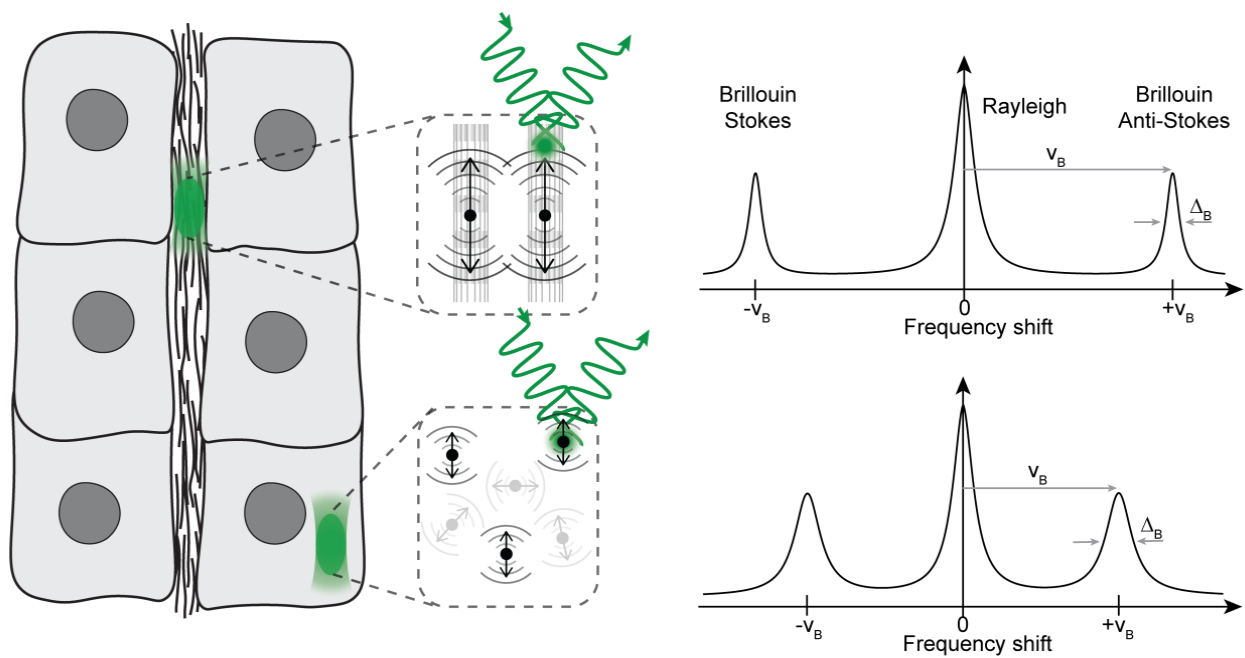

**Figure 1 | Brillouin scattering in heterogeneous biological samples.** Monochromatic laser light scatters from acoustic waves propagating in the longitudinal (axial) directions, giving rise to the Brillouin spectrum. In general, light scattered of solid components (e.g. collagen fibers – top) typically experiences a high Brillouin frequency shift $v_B$. In contrast, the spectrum from the liquid-like cytosol (bottom) results in lower shift, indicating a less rigid material, and larger linewidth $\Delta_B$, indicating a more viscous medium.

These sound waves exhibit an intrinsic dependence on the viscoelastic properties of the material, in particular its complex longitudinal modulus $M$. Light scattered elastically from a sample has the same frequency as the illumination (so-called Rayleigh light). A small fraction (~$10^{-12}$) of the incident light, however, can interact with the acoustic waves, and exchange energy and momentum in the process. This can intuitively be understood as the density (acoustic) wave acting as a grating which diffracts the light: as the grating is travelling with velocity *V*, the scattered light experiences a frequency shift due to the Doppler effect. This gives rise to two additional peaks in the scattered light spectrum (see **Fig. 1** and **Box 2**). Analysis of the Brillouin frequency shift can therefore provide, for a known material density and refractive index, a unique characterization of the material mechanical properties in terms of the real part *M'* of the longitudinal modulus. Furthermore, from the linewidth of the Brillouin peaks, the acoustic wave attenuation and hence viscosity *η* can be obtained through the imaginary part of the longitudinal modulus (loss modulus *M''*) (see **Box 2**). However, care must be taken to appropriately interpret the Brillouin spectra recorded from heterogeneous biological matter by considering the relevant spatial and temporal scales of the Brillouin scattering in relation to the biological process of interest (see **Box 3**).

---

**Box 2 | Relation of Brillouin spectrum to viscoelastic properties**

Brillouin spectroscopy yields information on the complex and frequency-dependent longitudinal modulus $M(v) = M'(v) + iM''(v)$. Here, the frequency $(v)$ dependence of the modulus highlights the fact that, e.g. in liquids and polymers, during the fast timescale of the material's deformation incurred by the acoustic wave, some of the slower molecular relaxation processes do not have any contribution, thus effectively "stiffening" up the material. t is worth noting that the longitudinal modulus is a different quantity with respect to the Young's modulus E: their definitions imply much higher (GPa) values of M compared to E (kPa) (see **Box 4**).

In the spectrum, the Brillouin peak's position is determined by the real part of *M* (*'storage modulus'* - see Eq. 1 and **Fig. 1**) which accounts for the elastic behavior, i.e. the stored elastic energy inside a sample. In particular, this Brillouin frequency shift, $v_B$, is typically on the order of 1-20 GHz and given by:

$$v_B = Vq. \qquad (1)$$

Here *q* is the exchanged wavevector given by:

$$q = \frac{2n}{\lambda_0} \sin(\theta/2) \qquad (2)$$

where *n* is the material refractive index, $\lambda_0$ is the incident wavelength, θ is the angle between the incident and scattered light, and *V* is the medium's acoustic velocity which in turn is related to the material density ρ and the real part of the complex longitudinal modulus (*M'*) through $V = \sqrt{M'/\rho}$ [55].

The imaginary part *M"* (*'loss modulus'*) on the other hand bears information about a sample's acoustic attenuation and thus viscous-like properties. This is because many materials, biological media included, will also dissipate the elastic energy of the travelling density waves, and therefore lead to the Brillouin side bands acquiring a natural line width $\Delta_B$ given by[55]:

$$\Delta_B = \Gamma q^2 \qquad (3)$$

where $\Gamma$ is the attenuation coefficient. From this the viscosity $\eta$ can be inferred through $\eta = 2\Gamma\rho$, where $\rho$ is the mass density, and the loss modulus via $M''=2\pi\nu\,\eta$. Care must be taken, however, to account for spurious effects that distort and broaden the measured linewidth and that are not related to $\Gamma$, such as the contributions from high-NA geometries (a large distribution of possible scattering angle's θ in Equations 1-3)[24], the finite width (resolution) of the spectrometer, as well as effects from multiple scattering. Most of these can however be independently measured or modelled and thus used to appropriately de-convolve the raw spectrum[52,56].

The longitudinal modulus is also related to other moduli of elasticity, in particular the bulk $K$ and shear $G$ moduli as:

$$M = K + \frac{4}{3}G \qquad (4)$$

The bulk modulus $K$ is a measure of the compressibility of a material and defined as the ratio of a stress radially applied to the resulting relative change in volume[57]. For highly incompressible materials such as liquids and solids (and most likely biological matter), $M$ and $K$ will be relatively large and almost equivalent. The shear modulus $G$, on the contrary, is defined as the ratio of shear stress to shear deformation at constant volume. It has a much smaller value than $K$, unless the shear stress is applied fast enough to prevent the system from relaxing, in which case, $G$ and $K$ assume similar (GPa) values (also see **Fig. 3** and **Box 4**).

---

**Box 3 | Spatial and temporal scales in Brillouin scattering**

Biological samples are highly heterogeneous media on many different time and length scales, which can profoundly affect the propagation and dissipation of acoustic waves and hence influence the measured spectrum dependent on the mechanical homogeneity of the biological structure of interest (see **Fig. 2**).

*Acoustic wavelength scale ($\mathcal{L}_A$):* The smallest relevant length scale pertains to the size of the probed acoustic waves inside the sample, given by $\mathcal{L}_A = \lambda/2n$ (e.g. ~200nm), where λ is the wavelength of the incident light and $n$ is the typical refractive index of tissue. This is typically on the order of ~200nm (for $\lambda$~550nm and $n$ ~1.35).

*Light collection scale ($\mathcal{L}_{PSF}$):* This is set by the spatial extent of the optical focus and thus determines the effective scattering volume. Practically this length scale is set by the point-spread function (PSF) of the microscope which depends e.g. on the microscope objective's NA and the pinhole diameter in a confocal setting. Typically, the size of the PSF, $\mathcal{L}_{PSF}$, is between 1-5μm in the axial direction in which the acoustic modes are probed.

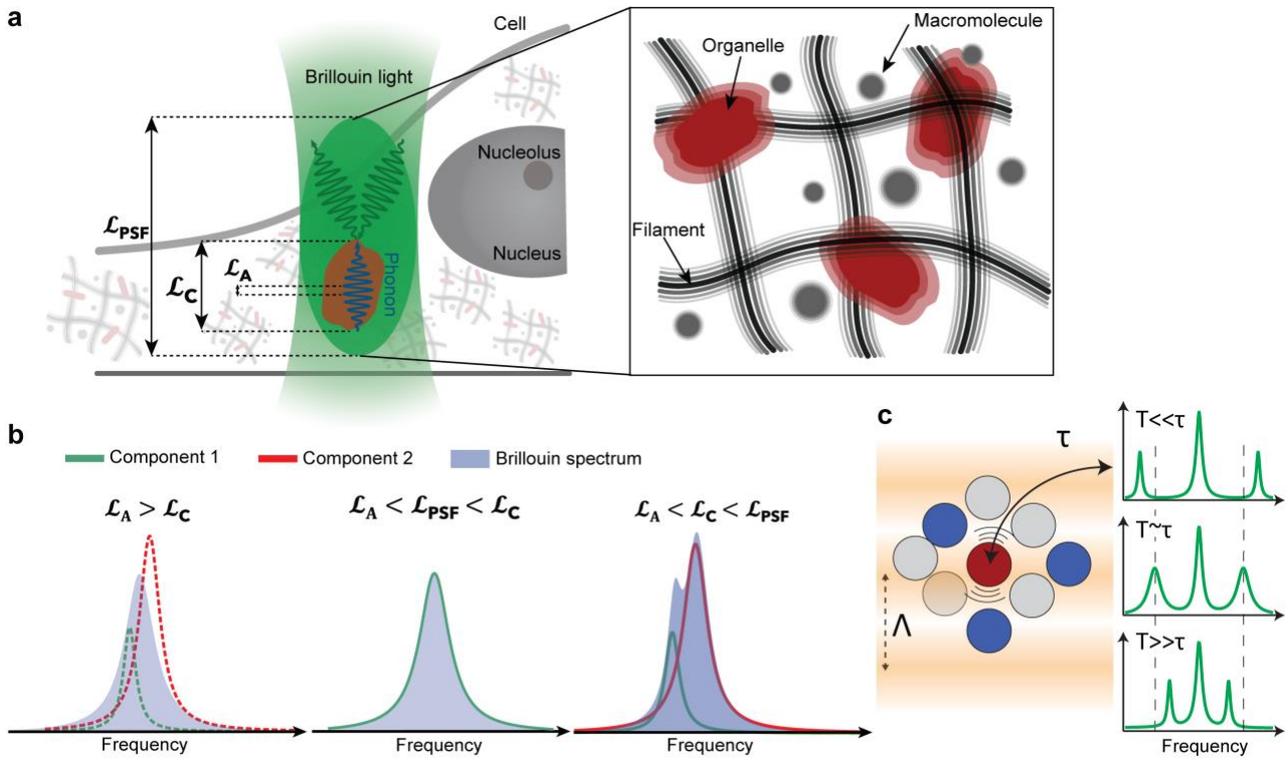

**Figure 2 | Schematic of Brillouin scattering**. **(a)** Visible (green) light interacts with acoustic wave (phonon, blue) within the scattering volume through Brillouin scattering. $\mathcal{L}_A, \mathcal{L}_C$ and $\mathcal{L}_{PSF}$ denote the relevant length scales for this interaction, given by acoustic wavelength, the size of the phonon supporting component and the PSF (scattering) volume, respectively. **(b)** Depending on the relative size of the biological structures in relation to the acoustic wavelength, individual components give rise to either a sum (left), a unique (middle) or an average (right) of their respective spectra. **(c)** The internal relaxation time of the molecules, τ, in relation to the acoustic wave's period $T = \Lambda/V$ can further influence the measured spectrum both in terms of frequency shift and linewidth.

*Component scale* ($\mathcal{L}_C$): This mark the typical size of a structure of homogeneous mechanical property (e.g. organelle, filament, etc.), whose length scale in relation to $\mathcal{L}_{PSF}$ and $\mathcal{L}_A$ is important. For sample structures that are smaller than $\mathcal{L}_A$, i.e. $\mathcal{L}_A \gg \mathcal{L}_C$, the acoustic field experiences an effective homogeneous medium and the response will therefore depend on the average elastic properties of the constituents within this length scale, and thus cannot be resolved individually[58,59] (**Fig. 2b**, left).

For larger structures, i.e. where $\mathcal{L}_C \gg \mathcal{L}_A$, two cases are possible:

*i)* $\mathcal{L}_{PSF} < \mathcal{L}_C$, this is the most desirable situation, in which the optical resolution is high enough so that the mechanical properties of a single component can the observed (**Fig. 2b**, middle).

*ii)* $\mathcal{L}_{PSF} > \mathcal{L}_C$, in this case several mechanical components occupy the same scattering volume, and the resulting Brillouin spectrum is the incoherent sum of the individual peaks originating from the different components. Therefore, one can expect multiple Brillouin peaks if the component's shifts are sufficiently different (**Fig. 2b**, right).

Temporal scales in Brillouin microscopy relate to the intrinsic molecular relaxation time within the sample, as well as to more practical scales related to microscopy image formation. The former time scale, $\tau$, is the time required by the medium to *"rearrange"* its constituent particles (atoms and/or molecules) upon perturbation by the acoustic field (see **Fig.2c**). The actual value of $\tau$ depends on material properties such as temperature, composition, physical-chemical state, etc. and can ranges from few picoseconds (e.g. water at room temperature), to milliseconds (highly viscous liquids), to seconds (glass-like bio-matter like stress granules or amyloid plaques). If the system is supporting sound waves with period $T \gg \tau$, the medium has sufficient time to rearrange itself during the density perturbations associated with the acoustic wave that take place on the timescale $T= \Lambda/V$ (practically on the order of ~50-500ps). The travelling wave then sees a relaxed medium, which for any local density change finds its best energetic arrangement. If on the contrary the period is much faster than the relaxation time, $T \ll \tau$, the medium cannot arrange itself to follow the density fluctuation, and it appears "quenched", i.e. more rigid. In this case, all the elastic moduli are larger than in the former: the sound velocity is larger at high than at low frequency. It is also relevant to note that liquids can support propagating shear waves at frequencies higher than $1/\tau$. More interesting is the case when $T$ is close to $\tau$. Here, the sound wave can exchange energy with the medium, which leads to an enhanced dissipation and thus a higher Brillouin peaks linewidth (see **Fig.2c**).

On the instrumentation side, Brillouin signal acquisition typically requires integration of photons on the detector spanning several milliseconds to seconds per image voxel. Therefore, the recorded spectrum must essentially be treated as an average over this time-scale. Importantly, intracellular processes such as protein diffusion and aggregation, as well as cytoskeletal network turn-over rates, can therefore influence the locally probed elastic properties. Finally, the rather long integration time per image voxel entails rather slow scanning speeds, therefore acquiring an entire Brillouin map can take on the order of several minutes to hours. This time scale also has to be accounted for in relation to the (dynamic) biological processes under investigation.

**Which (bio-)mechanical properties can be measured?**

As described above, Brillouin measurements probe the complex longitudinal modulus. In a biological material, the real part of the longitudinal modulus and hence the frequency shift, $v_B$, depends on the intrinsic properties of individual components in the probed region, their length-scale, their level of cross-linking, the compressibility of the local microenvironment and the solid-liquid volume fraction[60]. At the intracellular level, elasticity is governed by the cell's complex internal structure, of which the cytoskeleton and the liquid fraction are understood to be paramount. Actin filaments, microtubules, and intermediate filaments contribute to overall elasticity to varying degrees in different cell types[61]. Specifically, their filament length and the level of crosslinking have been shown to lead to networks with different mechanical properties[29]. Moreover, cell stiffness also varies due to water efflux as this affects intracellular molecular crowding[62]. Thus, in some cell types, it has been suggested that the cytoplasm can be described as a poroelastic material[52], i.e. a porous elastic meshwork of (inter-connected) solid components (cytoskeleton, organelles, macromolecules) bathed in an interstitial fluid (cytosol). This framework has recently been extended to describe the structure of and metastatic potential of tumors[64]. In the extracellular environment, extracellular matrices can also be described as elastic networks embedded in extracellular fluid. At a tissue level, cell adhesion might lead to coupling of several solid mesh-works and thus generate a larger effective network[64]. Molecular mechanisms have been identified that allow cells to regulate mechanical properties such as network cross-linking[65] and branching[66], variations in the solid-liquid volume fraction[67] and pre-stress in their cytoskeletal[68] and ECM networks[69], thus raising the question to which extent these can be observed in the Brillouin spectrum of cells and tissues.

In fact, the frequency shift in Brillouin scattering has been shown to be sensitive to two major mechanisms of cytoskeleton stiffening: actin polymerization and branching of actin fibers[40,70]. Moreover, protein aggregates and liquid phase separated organelles can also lead to a differential Brillouin shift, as observed in amyloid plaques[52], nucleoli[41] and stress granules[43]. Last, Brillouin shift has been shown to be sensitive to myosin contractility during *Nematostella* development[71]. At the extracellular level, an increase in Brillouin shift has been observed in polyacrylamide gels of increasing rigidity[70]. In contrast, the biological origin of the imaginary part of the longitudinal modulus and hence the viscosity $\eta$, has not been systematically addressed. From a physical point of view, viscous properties can provide information on the physical state of matter. Thus, variations in Brillouin linewidth can in turn yield fundamental insights on critical liquid-to-solid phase transitions occurring in intracellular compartments, such as stress granules, in response to protein expression or disease progression[43]. At a cellular level, the presence of a lipid-rich layer surrounding amyloid plaques, as well as the level of hydration, have been shown to influence the line-width[52]. At the tissue level, the level of cell-cell adhesion has been associated with a modified viscosity assessed through an increase in Brillouin line-width in cellular aggregates[64]. However, for any linewidth interpretations, it is important to discriminate viscous effects from other factors that lead to a linewidth broadening, such as e.g. a heterogeneous mixture of compartments within the probed volume (see **Box 3**).

All the proof of principle experiments performed so far have demonstrated that the Brillouin shift is sensitive to filament crosslinking and the solid-liquid volume fraction. However, more systematic studies involving different cell types and controlled, specific perturbations will be needed to elucidate what networks characteristics (such as length-scale of their effective components or average crosslinking distance) can be sensitive to acoustic fields and thus contribute to the Brillouin spectrum. Moreover, future experiments will also have to show what cellular and tissue components and properties affect the line-width and thus contain information on the viscosity.

**Technical state-of-the-art**

In Brillouin microscopy, the sample is illuminated by a narrow-band frequency laser source and the scattered light is spectrally analyzed by a high-resolution (sub-GHz) spectrometer. This is often done in backscattering (epi-) geometries to avoid complex spectral broadening[56] and to allow the use of high NA objectives. Traditionally, Fabry–Perot interferometers were involved in spectrometer designs, typically involving a scanning multi-pass configuration[72]. While such design yield high spectral resolution and contrast, their low throughput (high loss) and long spectral scan times made applications in biology challenging. This was overcome by Scarcelli and Yun in 2008[39] by developing a high resolution spectrometer based on a so-called virtually imaged phased array (VIPA)[73]. The VIPA is essentially a modified version of a Fabry-Perot etalon capable of spatially dispersing the spectrum without the need for scanning the optical cavity of the interferometer, which enables convenient and efficient spectrum acquisition with a scientific camera. Further instrument development efforts have since significantly improved the performance of VIPA based devices. Different methods based on multi-stage schematics[74,75], destructive interference[76], etalon filtering[77], cell absorption[78] and dark-field illumination[79] have been proposed and realized in the attempt of further suppressing the elastic background signal and thus to improve the contrast when imaging in biological media, for which extinction ratios of ~70-90dB are required to achieve shot-noise limited detection . Furthermore, approaches exploiting beam apodization[40,80], Lyot filtering[81] and diffraction masks[43], have been demonstrated to increase the intrinsic spectral contrast of traditional VIPA spectrometer without compromising their throughput efficiency. This is of utmost importance, as realistic Brillouin signals obtained with a few mW of incident light are only composed of ~$10^4$ photons/sec. Moreover, Brillouin spectrometers have been coupled with existing methods such as Raman[82,83] and fluorescence imaging[42] in order to correlate the probed elasticity with morphological and structural as well as biochemical properties. To capture biologically relevant changes in material properties, often a measurement precision of <10MHz is needed and achieved through diligent spectral calibration and analysis. On this end, methods based on spectral autocorrelation analysis have been shown to improve the precision of the Brillouin peak localization[84], although this does not provide information about the Brillouin peak linewidth. Altogether, these recent advancements have markedly expanded the Brillouin 'toolbox' and thus made routine applications to current questions in biology and medicine possible.

**Limitations and challenges**

Despite these advancements, Brillouin scattering also comes with certain limitations that need to be acknowledged. To properly quantify the real and imaginary parts of the longitudinal modulus, an exact knowledge of the local refractive index and material density is necessary by definition. As such measurements[85] are experimentally challenging to achieve *in-situ*, this is often stated as an intrinsic limitation of Brillouin microscopy. Yet, the Lorentz-Lorenz relation predicts that that the refractive index squared ($n^2$) scales with the mass density ($\rho$), such that variations in these parameters in a heterogeneous sample will, to a good approximation, cancel each other out. Moreover, for biological investigations, quantitative material properties are often not truly required unless to inform mathematical models and numerical simulations, and relative changes between experiments can already provide valuable mechanical insights. Nevertheless, the ongoing quest to combine current Brillouin imaging systems with existing techniques that can measure refractive index or density *in-situ*, such as tomographic phase microscopy[86] or digital holographic microscopy[87], seem promising. But even in absence of these, other approaches can be taken to understand and single out the role of mechanical properties from the Brillouin measurements. Here, the loss tangent[64], $\tan \varphi = M''/M' = 4\pi \Delta_B/\nu_B$, does not depend on the refractive index $n$ (or mass density $\rho$) and thus provides a simple approach to determine whether mechanical properties are the main contributor to the observed changes, e.g. across time.

Another intrinsic disadvantage of Brillouin scattering is the weakness of the measured signal, which entails relatively long data acquisition times and potentially harmful illumination dosages. This has often limited Brillouin microscopy to *ex-vivo* samples or relatively static biological conditions. While more red-shifted illumination can partly mitigate phototoxicity effects, significant instrument development will be necessary to increase the acquisition speed, in turn enabling more wide-spread *in-vivo* experimentation and to potentially aid in the development of novel clinical applications. In this respect, nonlinear stimulated[88] or impulsive[89] Brillouin modalities and line-scanning[90] approaches have shown promising progress for rapid three-dimensional mechanical imaging, although more work will be needed to turn them into truly live-imaging modalities.

On the analysis side, careful interpretation of the obtained Brillouin spectra is needed to properly put the mechanical parameters into context with previous research in the field (see **Box 1,4**). Recently, Wu *et al.*[53] highlighted the fact that for highly hydrated materials, such as hydrogels, the Brillouin shift, and hence the real part of the longitudinal modulus, does not correlate with the Young's modulus. This argues against a straightforward interpretation of the Brillouin shift in terms of '*stiffness*', and indeed judicious interpretation is required to link Brillouin signals to underlying structural and mechanical processes. Yet, Wu *et al.* investigated hydrogels of artificially large (>90%) water content, significantly different from realistic conditions encountered in typical cells and tissues (~60-80%). While the Brillouin frequency shift is indeed more sensitive to water content also in these conditions[54], state-of-the-art instruments still provide enough sensitivity to detect solid-part compressibility, provided the water content does not change significantly. Indeed, the longitudinal modulus and Young's modulus are not directly related to one-another, and thus we caution the reader from considering them both the same proxy for *'stiffness'* (see **Box 4**). While phenomenological correlations between them may often exist, they need to be established through careful calibration and in a sample-dependent manner.

**Box 4 | The longitudinal modulus in Brillouin microscopy in relation to other moduli**

It is important to realize that Brillouin scattering probes the elasticity via a different physical process with respect to other mechanical measurements done e.g. with an AFM (**Fig. 3**). The main difference is that the longitudinal modulus $M$ probes the ratio of uniaxial stress to strain in a confined condition, i.e. in which the material is not allowed to expand sideways, thereby changing its density and/or volume. The '*stiffness*' measured by Brillouin scattering is therefore fundamentally different from the often used tensile (Young's) modulus $E$, which requires the volume to be kept constant. As a consequence, although both $M$ and $E$ share the same units (Pa), the longitudinal modulus probed in Brillouin scattering is in general much higher (~GPa).

Formally, the longitudinal modulus is related to the Young's modulus, $E$, by $M = E(1 – v)/[(1 + v)(1 – 2v)]$, where $v$ is Poisson's ratio. However, this gives the often-wrong impression that $M$ and $E$ are "proportional" to each other. While indeed $v$ is very close to 0.5 in biological materials and thus explains the few orders of magnitude difference between $M$ than $E$, it is important to emphasize that a universal relationship does not exist. In particular, the Poisson ratio bears a strong frequency dependence. Nevertheless, empirical relationships between $M$ and $E$ have been established[40,91] for specific cell types and experimental conditions.

In view of these differences among different mechanical measurements, care must be taken when attempting to compare these fundamentally different, yet complementary, mechanical parameters in the context of a straightforward '*stiffness*' measure.

**Future potential of Brillouin microscopy in biology and medicine**

From a biological and medical standpoint Brillouin microscopy opens the door to several exciting avenues both in basic research and for early diagnosis (see **Fig. 4**). The ability to image mechanical differences inside tissues could crucially contribute to deciphering how tissue elasticity and viscosity contribute to animal development, organogenesis and disease progression, where the interplay between mechanical properties, forces and signaling determines size and shape (see **Fig. 4b,c**). In particular, the combination of Brillouin microscopy with specific molecular perturbations could reveal, for example, to what extent adhesion or cortical tension contribute to tissue viscoelasticity, which are crucial for morphogenesis. Moreover, at the cellular level, we envision that Brillouin microscopy could be combined with single-cell sequencing methods to identify unique molecular fingerprints that control cellular viscoelasticity. Last, in the case where fibril networks can sustain acoustic modes, Brillouin microscopy could allow visualizing alignment and testing the role of cytoskeletal and ECM components *in-vivo*[92] in a plethora of biological questions[7,11]. It will be exciting to see to which level Brillouin microscopy can contribute to our understanding of, for example, cell motility, cell division or organ elongation, where fiber alignment and crosslinking have been shown to be paramount[93].

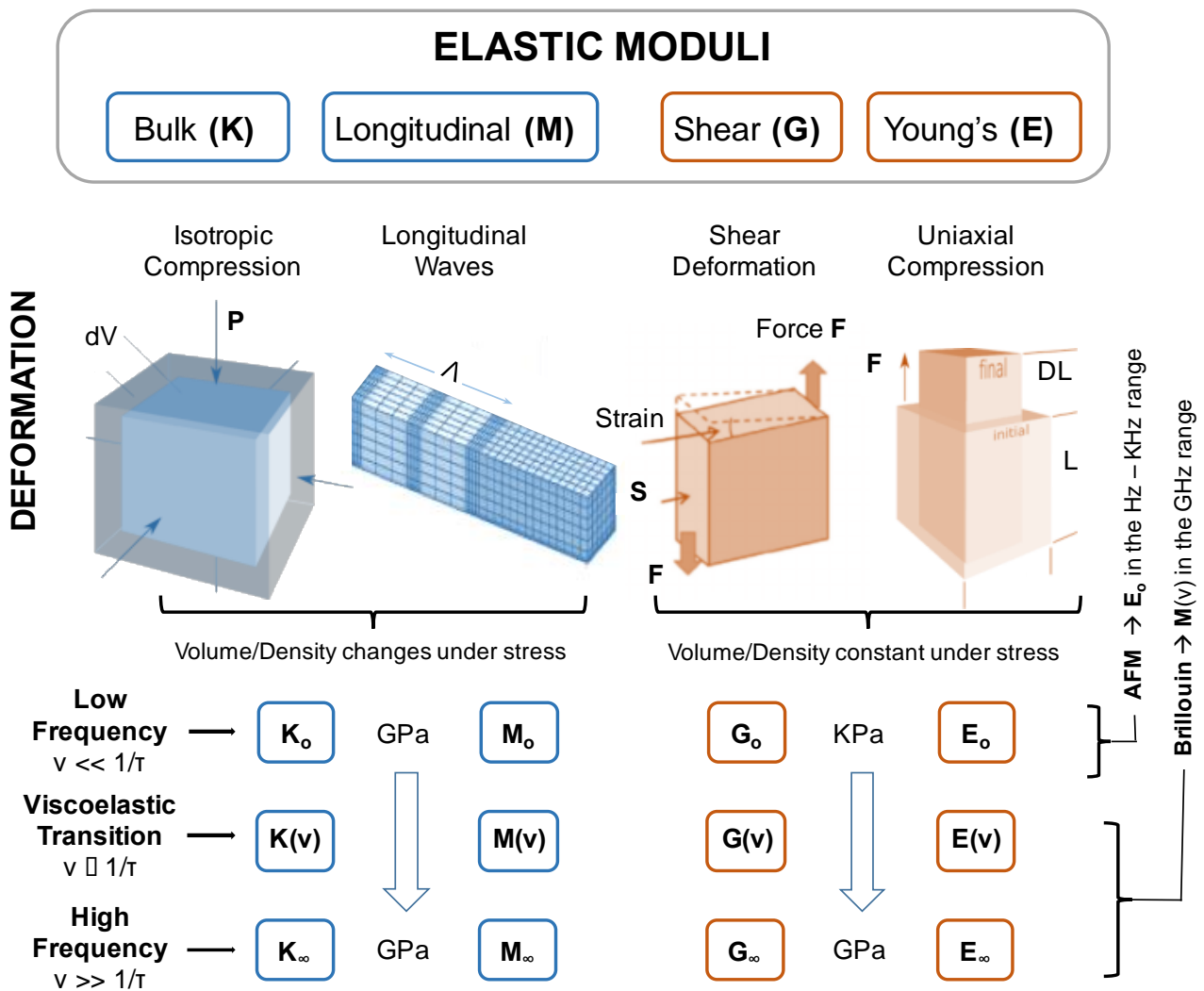

**Figure 3 | Synopsis of the main elastic moduli involved in mechanobiology.** The bulk (K) and longitudinal (M) moduli (blue) are associated with deformations that involve a change in the sample volume. On the other hand, the shear (G) and Young (E) moduli (red) refer to volume conserving deformations. As a consequence, in biological matter similarly to other liquids, gels or colloidal suspensions, K and M assume values in the GPa range at any frequencies, while G and E are almost vanishing (kPa range) for quasi-static deformations (low frequencies) where the system reacts as a liquid. At frequencies higher than those involved in the viscoelastic transitions ($\nu > \tau$), the bio-matter becomes solid-like, thereby G and E assume the typical values of a solid (GPa).

Furthermore, Brillouin microscopy has attracted interest also in the medical field as a tool for early diagnosis and to study the evolution of certain diseases. First clinical trials using Brillouin microscopy as a staging tool during the development of keratoconus have recently started (see **Fig. 4d**). Other eye pathologies that result in mechanical changes, like glaucoma, could also benefit from the development of ocular Brillouin devices. Additional pathologies that might benefit from Brillouin microscopy are cardiovascular diseases[48] and several cancer types[94,95] as mechanical differences have already been shown to be relevant for prognosis and overall survival[14]. In fact, efforts towards developing diagnostic blood vessel catheterization and endoscopic Brillouin tools are underway as they may enable early disease diagnosis.

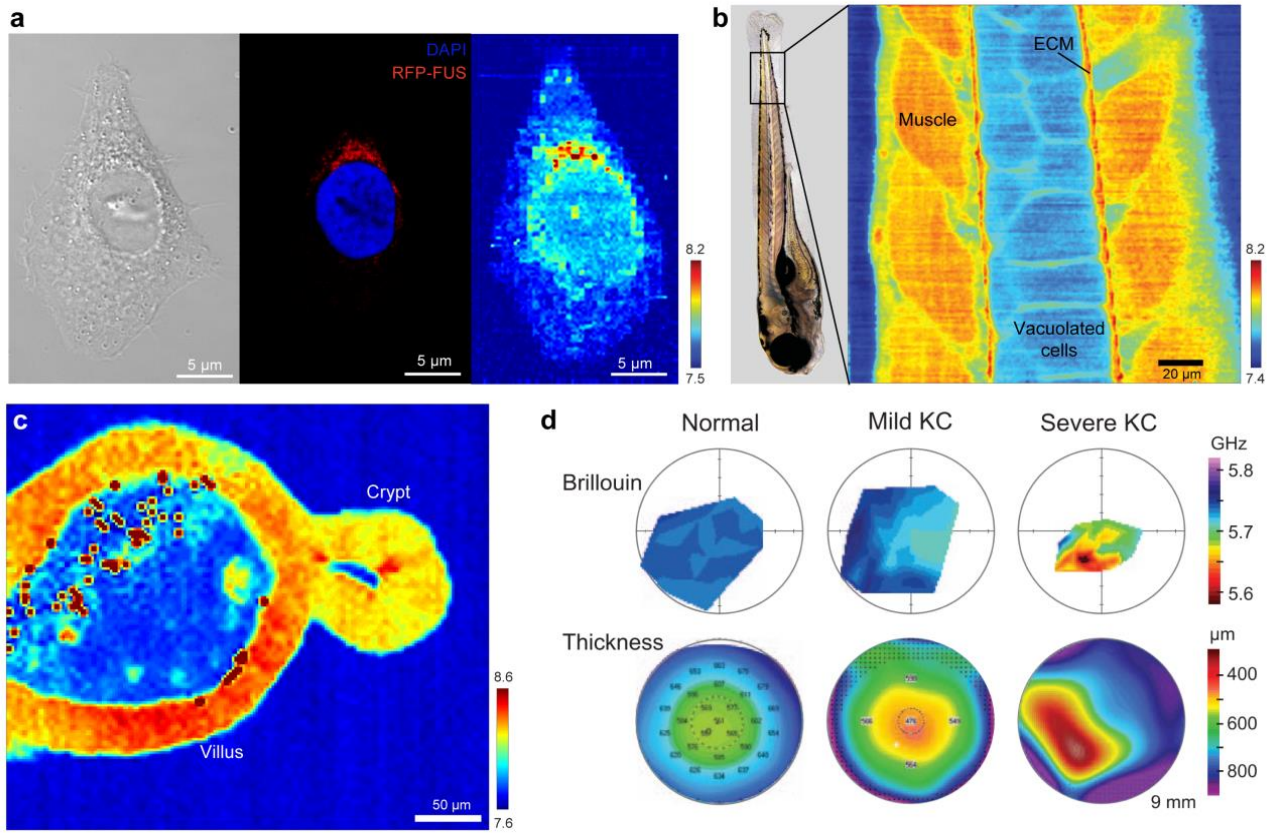

**Figure 4 | Brillouin microscopy in current biology and medicine.**
**(a)** High resolution Brillouin microscopy enables 3D mapping of the intracellular biomechanical properties in whole cells. DIC (left), immunofluorescence (middle) and Brillouin shift images (right) of HeLa cell line RFP-FUS$^{P525L}$ are shown for comparison. The red RFP signal shows the recruitment of RFP-FUS$^{P525L}$, present only in doxycycline-induced cells. Remarkably, the Brillouin image shows a higher frequency shift in response to the recruitment of ALS-linked RFP-FUS protein that occurs in cellular stress granules[43]. The nucleus and intranuclear nucleolus are also clearly distinguishable. **(b)** Brillouin shift image of live zebrafish tail tissue. Anatomical landmarks such as muscle, notochord vacuolated cells, and extracellular matrix show clear mechanical contrast[92]. **(c)** Brillouin shift map of a 5-day Matrigel cultured intestinal organoid with a crypt. The variations of average mechanics between different functional regions (crypt vs. villus) link tissue mechanics with crypt morphogenesis. **(d)** Brillouin measurement of keratoconus patients. Brillouin frequency shift (top) and corresponding pachymetry (bottom; Pentacam; Oculus Gmbh) maps of patients diagnosed with mild KC (stage 1; middle) and severe KC (right), in comparison to a normal subject (left). From Ref.[13]. Color bars denote Brillouin frequency shift in GHz in all panels.

**Outlook**

The field of mechanobiology has grown significantly in the last decade but the time and length scales where cell mechanics play a significant biological role are still debated. Arguably, the physical origin of *stiffness* has been conventionally associated with the Young's modulus measured by AFM. However, the emergence of Brillouin microscopy in biology raises the question whether this material property may be further or even better described by the longitudinal modulus. The integration of mechanical inputs differs in different biological and medical contexts, thus making it difficult to decipher what the most relevant quantitative method is. Going forward, we think that comparative studies involving viscoelastic frequency scaling in a range of biological samples (eg. cell types) and specific molecular perturbations will be critical to gain a better understanding of the origin and function of the longitudinal modulus inside biological tissues. This should of course go hand-in-hand with theoretical modelling taking molecular dynamics and structural parameters into account. Moreover, systematic assessment of the same biological process with largely varying microscope integration times could reveal separate molecular origins of the same Brillouin spectra. Altogether, such experiments could help to consolidate results obtained by Brillouin microscopy with other work in the field of mechanobiology. On the instrumentation side, advancements in the near future could involve employing synchrotron X-ray radiation (wavelengths 0.1 – 1 nm) to measure the properties of sound waves, and thus the biomechanics of cells and tissues, in the THz range. Beside the extended dynamical range, this would allow a much higher spatial resolution due to the much higher focusing capability of synchrotron radiation well below 100 nm.

Although a full understanding of the underlying complicated mechanics of biological constituents at high frequencies remains challenging, Brillouin microscopy has already provided fundamentally new insights by showing sensitivity to many mechanical processes that are biologically relevant. The unique abilities of Brillouin microscopy to measure the spatial and temporal modulations of mechanical properties within intact cells and tissues implies that this technology could have a profound impact on our understanding of biology and disease. Future advancements in instrument development and biophysical characterization across model systems will show whether and how these potentials can be fulfilled.

## Acknowledgements

We thank G. Scarcelli, C.C. Chan and K. Elsayad for insightful discussions and feedback on the manuscript, Q, Yang and P. Liberali for providing the samples shown in Fig. 4c, and C. Bevilacqua and M. Bergert for help with figures. This work was supported by the European Molecular Biology Laboratory (R.P., A.D.-M.), and the Deutsche Forschungsgemeinschaft (DFG) research grant DI 2205/2-1 (A.D.-M.).